\documentclass[usegraphicx]{mn2e}
\begin{document}

\title[Universal FJ relation]{The Universal Faber-Jackson Relation}

\author[R.H. Sanders] {R.H.~Sanders\\Kapteyn Astronomical Institute,
P.O.~Box 800,  9700 AV Groningen, The Netherlands}

 \date{received: ; accepted: }

\maketitle

\begin{abstract}
In the context of modified Newtonian dynamics, the fundamental plane, as the
observational signature of the Newtonian virial theorem,
is defined by high surface brightness objects
that deviate from being purely isothermal:  the line-of-sight
velocity dispersion should slowly decline with radius as
observed in luminous elliptical galaxies.  All high
surface brightness objects (e.g. globular clusters, ultra-compact
dwarfs) will lie, more or less, on the fundamental plane defined
by elliptical galaxies,
but low surface brightness objects (dwarf spheroidals) would 
be expected to deviate from this relation.  This is borne
out by observations.  With MOND, the Faber-Jackson relation
($L\propto \sigma^4$), ranging from globular clusters
to clusters of galaxies and including both high and low surface
brightness objects, is the more fundamental and universal
scaling relation in spite of its larger scatter.  Faber-Jackson
reflects the presence of an additional dimensional constant
(the MOND acceleration $a_0$) in the structure equation.

\end{abstract}

\section{Introduction}

The direct
observational signature of the Newtonian virial theorem
in elliptical galaxies emerged two decades 
ago with the discovery of the ``Fundamental Plane" 
(Djorgovski \& Davis 1987, Dressler et al. 1987).
Assuming that ellipticals are reasonably homologous
Newtonian pressure supported
systems, then they will comprise a two-parameter set:
if luminosity is proportional to mass then the virial
theorem implies that $L\propto \mathit{R_{eff}}{\sigma^2}$ where
$\mathit{R_{eff}}$ is the effective radius and $\sigma$ is the
line-of-sight velocity dispersion. 
This is essentially the observed Fundamental Plane relation.
Although
the connection between the Fundamental Plane (hereafter
FP) and the virial theorem was originally obscured by the
fact that the observationally inferred exponents in the 
above relation were not precisely as expected, it is now 
established that
this deviation is primarily due to a systematic variation in the 
mass-to-light ratio of ellipticals: M/L appears to slowly
increase with luminosity (see Faber et al. 1987 for an early realization
of the relevance of the virial theorem).  That the FP is the observational
manifestation of the virial theorem has become indisputable with
the determination, via gravitational lensing, of the mass-based
FP, the ``more Fundamental Plane"(Bolton et al. 2007).  Here 
surface density, $\Sigma$, replaces surface brightness, $I$, in the equivalent FP relation, $\mathit{R_{eff}} \propto \sigma^2/I$, and
thereby removes the effects of systematic and random variations in M/L.

The striking
aspect of the traditional FP is the small scatter about the expected
relation.  Such a precise relation surely reflects the
facts that elliptical galaxies comprise a rather homologous
class of objects with a fairly isotropic velocity distribution
(at least in the inner regions) and that the random fluctuation in M/L is
relatively small.  This small fluctuation in dynamical
M/L is puzzling  
given the current view of a galaxy as a luminous baryonic
component immersed in a more massive and extensive dark
halo composed of non-baryonic particles.  Ellipticals are
presumably assembled over a Hubble timescale via mergers -- 
``wet" or ``dry", ``major" or ``minor" --
of smaller mass aggregates;  it is not evident that 
such a fixed proportion of dark and baryon matter within the
visible object should survive what must be a rather chaotic process.
 
The second scaling relation for elliptical galaxies was
actually discovered long before the FP -- that is the
relation between luminosity and line-of-sight velocity
dispersion of the form $L\propto \sigma^\alpha$ where $3\le\alpha\le 5$.  
This correlation was first described in
a qualitative way by Morgan \& Mayall (1957) who remarked that
``For progressively fainter Virgo cluster ellipticals the spectral
lines tend to become narrower, as if there were a line width-absolute
magnitude effect for the brighter members."  This effect was given
its definite and quantitive form by Faber \& Jackson (1976) and
has become known as the Faber-Jackson relation (hereafter FJ).
The FJ relation implies that ellipticals comprise a one parameter
family:  given the velocity dispersion, the luminosity
(or luminous mass) follows.  The observed FJ relation does have a much
larger dispersion about its mean than does the FP, and this  
has led to the often-heard assertion that the FP has superseded the
FJ -- that the FJ represents a non-orthogonal 
projection of the FP onto a lower
dimensional parameter space (Franx \& de Zeeuw 1991).  

While it is certainly true that the FP, because of its much smaller
scatter, is superior to the FJ as a distance indicator for elliptical
galaxies, it is not evident
that the FJ is a simple projection of the FP;  nothing like Faber-Jackson
is required by the virial relation.  FJ implies that the mass of a 
pressure-supported system is correlated to the velocity dispersion more-or-less
independently of the size of the object.  I have argued (Sanders 1994) 
that this same correlation extends to the great clusters of galaxies in the
form of the gas mass - temperature relation
($M_g\propto T^2$).  Indeed, the relation $M\propto \sigma^4$
appears to broadly apply to all near-isothermal pressure supported 
astronomical systems from globular star clusters to clusters of galaxies.

In the context of the standard CDM-based cosmology, this correlation
must arise from aspects of structure formation.  In
the standard scenario,
galaxies, or pre-galactic bound objects, form at a definite epoch which implies that there is, more or less, a characteristic
density for protogalactic objects.  The existence of a characteristic
density combined with the virial theorem yields a mass-velocity
dispersion relation for halos of the form $M\propto \sigma^3$ which
seems to be borne out by cosmological N-body simulations
(e.g., Frenk et al. 1988).
However, such a relatively shallow power law 
cannot be extrapolated from globular
cluster scale objects to clusters of galaxies.  In addition, globular
clusters and clusters of galaxies form at different epochs via
different processes; the emergence of this apparently 
universal correlation is not obviously implied.

Milgrom's modified Newtonian dynamics (MOND) does
give rise to a mass-velocity dispersion relation 
the form $M\propto \sigma^4$ as an aspect of existent dynamics
rather than the contingencies of structure formation (Milgrom 1983a).
This was an early prediction of the hypothesis (Milgrom 1983b).  
MOND does not, however, inevitably predict existence of the
Fundamental Plane;
in fact, one might ask if the FP as a manifestation of the
{\it Newtonian} virial theorem is consistent with MOND.  
Here I consider this question and discuss how the FP arises
in the context of MOND.  I review these two global
scaling relations and their extension to the wide class of
pressure supported, near isothermal objects, and 
I compare the universality of the Fundamental
Plane to that of Faber-Jackson.  I point out that objects which
define the FP should have high surface brightness (and thus
be essentially Newtonian) 
within an effective radius, and I highlight a MOND prediction that
such objects must deviate
from an isothermal state with an observed velocity dispersion that declines
with radius.  Further, very low surface brightness objects, such as the
dwarf spheroidal galaxies, might not be expected to lie on the
FP defined by ellipticals and globular clusters.  I compare
these expectations with the observations.

\section{Global scaling relations in the context of MOND}

One of the original motivations for MOND as an acceleration-based
modification of Newtonian dynamics was to subsume the 
Tully-Fisher relation for spiral galaxies (Sanders \& McGaugh 2002).  
This is the observed
correlation between the luminosity and asymptotic constant rotation velocity
in spiral galaxies, but has since been more accurately described 
as an exact relation between baryonic mass and  
rotation velocity (McGaugh, et al. 2000), $M\propto {V_{rot}}^4$.  
Milgrom (1983b) pointed out
that a similar correlation should exist for pressure supported
systems such as elliptical galaxies:  the total (baryonic) mass
is roughly proportional to the fourth power of the velocity dispersion,
but non-homology as well as variations in the degree of anisotropy
of the velocity field introduce much more scatter
than is evident in Tully-Fisher.

Milgrom (1984) explored this further in his paper on MONDian 
isothermal spheres.  Solving the spherically symmetric
Jeans equation with the MOND expression for the effective
gravitational acceleration he was able to draw several powerful conclusions:

1.  MOND isothermal spheres, unlike their Newtonian counterparts,
have finite mass.

2.  Asymptotically the density falls as power law,
$\rho(r) \propto r^{-\alpha_\infty}$ where  
$\alpha_\infty\approx 4$.

3.  The ratio of the total mass to the fourth power of the 
radial velocity dispersion, $\sigma_r$
depends primarily upon the anisotropy parameter 
$$\beta = 1-{\sigma_t}^2/{\sigma_r}^2 \eqno(1)$$ ($\sigma_t$ is
the tangential velocity dispersion); specifically,
$$M={(Ga_0)^{-1}} (\alpha_\infty-2\beta)^2{\sigma_r^4},\eqno(2)$$
where $a_0$ (approximately $10^{-8}$ cm/s$^2$)
is the fundamental new parameter of the theory.
This forms the basis for the Faber-Jackson relation.

4. For all ``deep MOND" objects 
(having internal acceleration everywhere 
less than the $a_0$) it is the case that
$M/\sigma^4 = 9/4(Ga_0)^{-1}$ independently
of $\beta$,
where $\sigma$ is the three dimensional velocity dispersion 
(as an aside
I mention that the deep MOND analytic solution for $\beta=1/2$ corresponds
exactly to the Hernquist (1991) model for spherical systems, a model which
projects to a surface density distribution similar to that observed
for elliptical galaxies).

None of these conclusions anticipated the Fundamental Plane because
it was not then fully appreciated that luminous elliptical galaxies are 
essentially Newtonian systems within an effective radius; whereas, MONDian
isothermal spheres are in transition between Newton and MOND at 
$\mathit{R_{eff}}$.  That is
to say, isothermal spheres are not good representations of actual
elliptical galaxies; they are too MONDian.  
We may characterize the degree to which
an object is Newtonian within the bright inner regions (within $\mathit{R_{eff}}$)
by the parameter $\eta =\mathit{g_{eff}}/a_0$ where $\mathit{g_{eff}}$ is the gravitational
acceleration at $\mathit{R_{eff}}$.  By this criterion $\eta \approx 0.7$ for
MONDian isotropic isothermal spheres whereas $\eta\approx 6$ typically 
for real ellipticals.  This means
actual luminous elliptical galaxies should show little evidence
for dark matter within an effective radius, as was borne out
when detailed kinematic data became available for several of these objects
(Romanowsky et al. 2003).  This result is completely consistent with
the expectations of MOND for such high surface brightness systems
(Milgrom \& Sanders 2003).

Viewed in terms of MOND, the Newtonian nature of elliptical galaxies
requires that these objects are not isothermal;
the velocity dispersion should be observed to decrease with radius.  
An implication
is that such systems should be consistent with the Newtonian virial
theorem within the effective radius; that is to say, they will
define a FP relation.  On the other hand, deep MOND (low surface
brightness) objects might be expected to deviate from this FP
while being generally consistent with the universal FJ relation.

\section{The MOND basis for the Fundamental Plane}

MONDian isotropic, isothermal spheres possess a
fixed relation between effective radius and central
line-of-sight velocity dispersion of the form
$\sigma^2 =  870 \mathit{R_{eff}} $ where $\sigma$ is in km/s
and $\mathit{R_{eff}}$ is in kpc. 
However, actual elliptical galaxies exhibit a fairly 
broad distribution 
by effective radius and velocity dispersion
(J{\o}gensen et al. 1995, Sanders 2000), and, for a given 
velocity dispersion, all are much more compact (smaller
$\mathit{R_{eff}}$) than the isothermal sphere.

In order to reproduce the observed distribution of elliptical galaxies 
by $\mathit{R_{eff}}$ and $\sigma$ in the context of MOND 
I previously considered non-isothermal
non-isotropic models: specifically, high order polytropic spheres
with an increasing radial orbit anisotropy in
the outer regions (Sanders 2000).  To be consistent with
the observed scatter in these two quantities it was necessary
to consider a range of polytropic
indices (n=12-16 where ${\sigma_r}^2 \propto \rho^{1/n}$) 
and anisotropy radii (scaled in terms
of the effective radius); i.e., the models must deviate from
strict homology.
Matching the observed joint distribution of $\mathit{R_{eff}}$
and $\sigma$ with this range of models, I found that the 
observed FP is reproduced. In particular, the implied
high surface brightness means that the models are essentially
Newtonian within an effective radius, and this 
yields a small scatter about the mean (virial) relation.  
A weak systematic increase of M/L with luminosity is
also required, but the ensemble provides a good representation of
the lensing (mass-based) FP discovered by Bolton et al. 
(see Sanders \& Land 2008 for a discussion of the lensing-
based FP in the context of MOND).  The FJ relation is also
present in this ensemble of polytropic models
but with much larger scatter because the relation is
more sensitive to the necessary deviations from homology.  

The reliance upon such
specific polytropic models somewhat obscures the dynamics
underlying the appearance of a FP in the context of MOND;
that is to say, the polytropic assumption is in no sense
necessary for matching the FP.
The necessary ingredient is a deviation from an isothermal
state.  This becomes more evident when considering
a specific model for the mass distribution in elliptical
galaxies, i.e., the spherically symmetric Jaffe model (Jaffe 1983) with
a radial dependence of density given by
$$\rho(r) = {{\rho_0{r_j}^4}\over{r^2(r+r_j)^2}} \eqno(3)$$
where $r_J$ is the characteristic length scale of the model
and $\rho_0 = M/(4\pi {r_j}^3)$ is the characteristic density
($M$ being the total mass).  The Jaffe
model projects very nearly into the empirically fitted surface
density of elliptical galaxies (de Vaucouleurs 1958) where
the $\mathit{R_{eff}} = 0.763r_j$.  

Given this density distribution, I
numerically solve the Jeans equation for the run of radial
velocity dispersion:
$${{d{\sigma_r}^2}\over{dr}} - {{d \log(\rho)}\over {d\log(r)}} \sigma^2
= -g\eqno(4)$$ where I have assumed that the velocity field is
isotropic ($\beta=0$).
Here $g$ is the true gravitational force which, in terms of MOND, is
given by $g\mu(|g|/a0) = g_n$ with $g_n$ being the usual Newtonian
force.  The interpolating function is taken to be $\mu(x) = x/(x^2+1)^{0.5}$
with $a_0 = 10^{-8}$ cms$^{-2}$ as implied by the rotation curves of spiral galaxies
(Bottema et al. 2001).

\begin{figure}
\resizebox{\hsize}{!}{\includegraphics{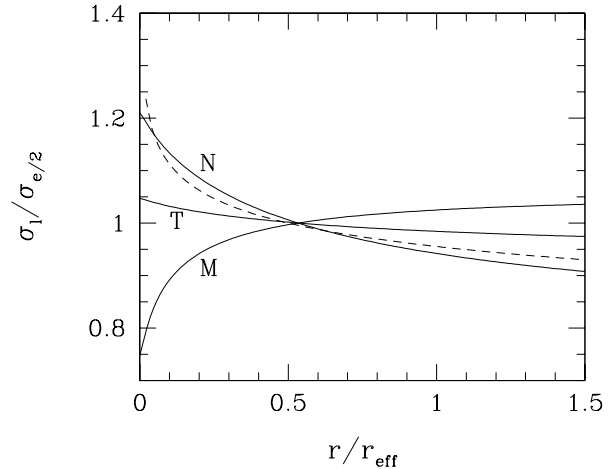}}
\caption[]{The intensity weighted line-of-sight velocity
dispersion within a circular aperture plotted as a function
of the circular aperture.  The velocity dispersion is given in
terms of that measured within one-half an effective radius and
the radius is given in terms of the effective radius.  The solid
curves show the run of $\sigma_l$ for three values of the
the parameter $\eta$ which is a measure of the internal acceleration
in the object.  The curve labeled N ($\eta = 7.8$) is a highly 
Newtonian object; that labelled T ($\eta = 0.9$) is in transition
between Newton and MOND; and curve M is a deep MOND object ($\eta=0.2$)
The dashed line shows a power law fit to the observed run of $\sigma_l$
in elliptical galaxies (Cappellari et al 2006).}
\label{}
\end{figure}

In Fig. 1 I show the results of such an integration for three different
values of the parameter $\eta = \mathit{g_{eff}}/a_0$.  This is the mean
intensity weighted line-of-sight
velocity dispersion within a circular aperture, $\sigma_l$, scaled
to its value within one-half an effective radius plotted as a function of
radius in units of the effective radius.  Three cases are shown:
$\eta = 7.8$ which is highly Newtonian within an effective radius
(labelled N); $\eta = 0.9$ which is in transition between Newton
and MOND at an effective radius (T); and $\eta = 0.2$ which is in the deep
MOND limit (M).  In all cases, for an isotropic velocity distribution,
the velocity dispersion approaches a constant value at large radii.

Fig.\ 1 illustrates the point that Newtonian objects
must deviate from isothermal with a velocity dispersion that declines within
an effective radius.  Transitional Jaffe models are almost isothermal;
the MOND isothermal sphere is such a transitional object.  Deep MOND
objects exhibit a velocity dispersion rising to an almost constant
value if the velocity field is isotropic; indeed, for small values of $\eta$
this form of the velocity dispersion profile becomes independent of
$\eta$.  This is all
reminiscent of the form of rotation curves in
high to low surface brightness disk galaxies.

We would expect objects which define the FP 
(i.e., consistent with the Newtonian virial theorem) to be
well into the Newtonian regime, i.e., $\eta >> 1$.  This is apparently
the case as is shown by the dashed curve in Fig.\ 1, which is a power
law fit by Cappellari et al. (2006) to the mean observed run of $\sigma_l$ ($\propto r^{-0.066}$) for 25 early type galaxies.  That is to say, these
observations are consistent with the MOND expectation of a declining
velocity dispersion profile for high surface brightness elliptical 
galaxies defining the FP. 

The galaxies in the sample of Cappellari et al. are plotted as
a fundamental plane relation in Fig.\ 2.  The crosses show
the luminosity vs. the virial quantity ${\sigma_e}^2 \mathit{R_{eff}}/G$
where $\sigma_e$ is the intensity weighted mean line-of-sight
velocity dispersion within an effective radius.  Units are 
$10^{11}$ L$_\odot$ and M$_\odot$.

The expectation from the virial theorem is
$$M = {c_1} {\sigma_e}^2 \mathit{R_{eff}}/G.\eqno(5)$$
where ${c_1}$ is the structure constant which depends
upon the MOND parameter $\eta$, or, alternatively, upon
the contribution of dark matter to the total mass within
an effective radius.  The two parallel lines in
Fig.\ 2 show this relation for two different values of $c_1$
For the upper curve, labelled $N$, the structure constant, $c_1=4.54$,
is that appropriate for the Newtonian Jaffe model ($\eta =10$) or an equivalent
dark matter fraction within an effective radius of about 0.1.
If all systems were represented 
perfectly by isotropic Newtonian Jaffe models with $M/L=1$ then they would lie
on this line.
For the lower line labelled $M$, the structure constant, $c_1 = 0.74$, corresponds
to a deep MOND Jaffe model ($\eta = 0.1$) or an equivalent dark matter
fraction within $\mathit{R_{eff}}$ of 0.9.  Note that deep
MOND Jaffe models with a fixed value of $\eta$ will also define 
a virial FP relation because they correspond to homologous objects with
a fixed fraction of dark matter within the effective radius.  Because
lower accelerations (lower $\eta$) corresponds to more dark matter
(higher effective M/L) the MOND line lies to the right of the
Newtonian line in this figure.

\begin{figure}
\resizebox{\hsize}{!}{\includegraphics{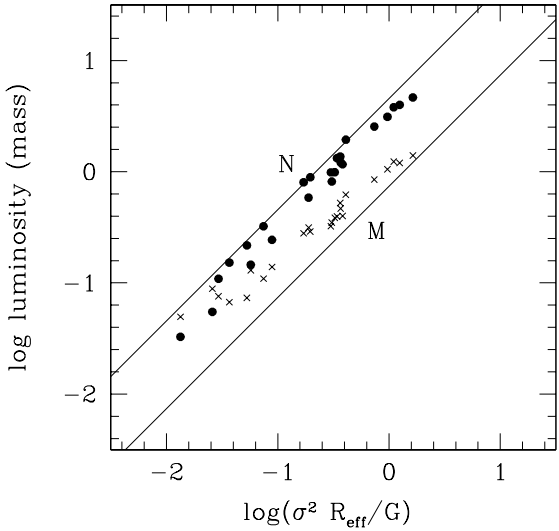}}
\caption{A log-log plot of luminosity vs. the virial
quantity ${\sigma}^2 \mathit{R_{eff}}/G$ for the early type
galaxies observed by Cappellari et al. (2006).  $\mathit{R_{eff}}$ is the half-light
radius and $\sigma$ is the line of sight velocity
dispersion within an effective radius.  Units are $10^{11}$ L$_\odot$ 
and M$\odot$.  The crosses are show the luminosity vs. the virial
quantity and the solid points are the same galaxies but with
mass, estimated from population synthesis M/L, plotted against
the virial quantity.
The parallel lines correspond to the virial quantity
$M = c_1{\sigma_e}^2\mathit{R_{eff}}/G$ where $\sigma_e$ is the intensity
weighted los velocity dispersion within an effective radius.
The upper line is appropriate
to a near Newtonian system ($\eta=10$) and the lower
line to deep MOND objects ($\eta = 0.1$), or, alternatively, to
homologous objects with a fixed fraction of dark matter within
$\mathit{R_{eff}}$
}
\label{}
\end{figure}

The solid points in Fig.\ 2 are the same systems but with
stellar {\it mass} plotted against the virial quantity.  The
masses are estimated from the luminosity multiplied by $M/L$ derived from 
population synthesis
models (Cappellari et al. 2006).  Now we see that the
points lie very near the virial relation for pure
Newtonian Jaffe models (small discrepancy or little dark
matter within an effective radius).  With MOND, this is the expected result
for high surface brightness galaxies.

There are other classes of high 
surface brightness objects with a radially declining velocity dispersion
which, apart from small shifts due to variations in homology, should
lie on the same FP: specifically, dwarf elliptical
galaxies, the ultra-compact dwarfs, and the globular star clusters.

\begin{figure}
\resizebox{\hsize}{!}{\includegraphics{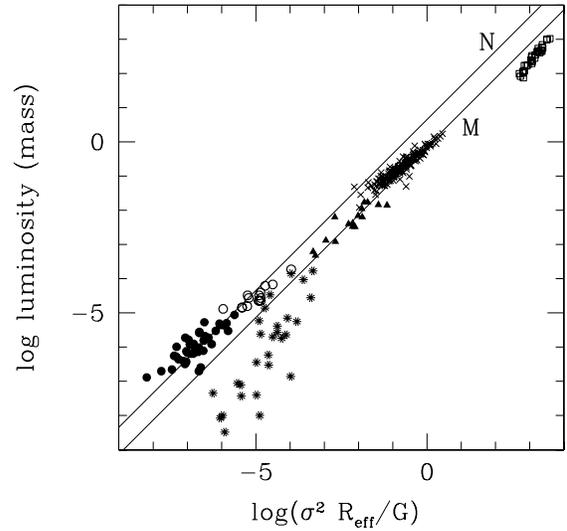}}
\caption{A log-log plot of luminosity vs. the virial
quantity ${\sigma}^2 \mathit{R_{eff}}/G$ for several classes of
hot pressure supported objects.  $\mathit{R_{eff}}$ is the half-light
radius and $\sigma$ is the line of sight velocity
dispersion.  Units are $10^{11}$ L$_\odot$ and M$_\odot$.
The solid points are globular star clusters
in the galactic halo (Pryor \& Meylan 1993, Trager et al. 1993),
the starred points are dwarf spheroidal galaxies (Walker et al.
2009, 2010),
the open points are the ultra-compact dwarfs, UCDs (Mieske et al. 2008),
the solid triangles are dwarf elliptical galaxies (Bender et al. 1992),
the crosses are luminous elliptical galaxies (J{\o}rgensen et al. 1996),
the open squares are X-ray emitting clusters of galaxies (Croston et al.
2008).  The data is quite heterogenous between the various classes
of objects:  luminosity is measured in different photometric
bands, and the velocity dispersion is, in some cases, the central
velocity dispersion within a fixed diaphragm (globular clusters), or
the intensity-weighted velocity dispersion within an
effective radius (dwarf ellipticals and UCDs), or the mean velocity
dispersion within a fixed linear radius (ellipticals).  For the
clusters it is the gas mass that is plotted on the vertical
axis.  As in Fig.\ 2 the parallel lines correspond to the virial quantity
$M = c_1{\sigma_l}^2\mathit{R_{eff}}/G$ where $\sigma$ is the intensity
weighted los velocity dispersion within an effective radius.
The upper line is appropriate
to a near Newtonian system ($\eta=10$) and the lower
line to deep MOND objects ($\eta = 0.1$), or, alternatively, to
homologous objects with a fixed fraction of dark matter within
$\mathit{R_{eff}}$
}
\label{}
\end{figure}

In Fig. 3 a wide range of objects is plotted on a virial
representation of the FP.  Here again the logarithm of the luminosity ($10^{11}$
$L_\odot$), a proxy for the mass,
is plotted against the logarithm of the virial quantity $\sigma^2 \mathit{R_{eff}}/G$
($10^{11}$ $M_\odot$).  If the objects are Newtonian (satisfying the
Newtonian virial theorem) and homologous with constant M/L they should lie on the FP defined by the elliptical galaxies.  
On this plot the ellipticals drawn from the
larger sample of J{\o}rgensen et al. (1996)
are shown as crosses; the globular
clusters are solid points, the triangles are dwarf elliptical; the
open points are ultra-compact dwarfs; and the stars are the dwarf
spheroidal companions of the Milky Way.  The open squares are
X-ray emitting clusters of galaxies but with gas mass plotted
rather than the luminosity. The references for these observations
are given in the figure caption.

Points significantly below the parallel lines have a large Newtonian
dynamical M/L, the conventional explanation being the presence 
of dark matter.  For example, the galaxy clusters are well 
below the Newtonian
line reflecting the classical virial discrepancy 
in these objects. 
 
The essential feature of this plot is that the high surface brightness
objects ranging from ellipticals to globular clusters define
a fairly narrow fundamental plane which is consistent with the virial theorem.
There are, of course, deviations due to non-homology 
and, more seriously, to the heterogeneity of the data samples. 
But the point is that the same FP is delineated by all
high surface brightness (i.e., high internal acceleration) objects.
\begin{figure}
\resizebox{\hsize}{!}{\includegraphics{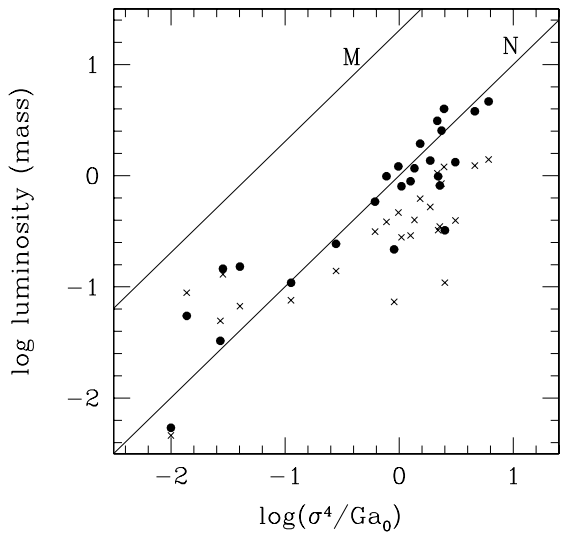}}
\caption[]{A log-log plot of the luminosity vs. the Faber-Jackson
quantity ${\sigma}^4/Ga_0$ in units of $10^{11}$ L$_\odot$ or
M$_\odot$ for the 25 early type galaxies observed by Cappellari
et al.   The parallel lines correspond to 
$M=c_2{\sigma_l}^4/Ga_0$ for the Jaffe model where again $\sigma_l$
is the intensity-weighted velocity dispersion within $\mathit{R_{eff}}$.  The upper
line is appropriate to a deep MOND object with $\eta=0.1$ (or to a
dark matter dominated homologous system) and the lower line to
a near Newtonian system with $\eta=10$ (or to a visible matter dominated
system).}
\label{}
\end{figure}

The low-surface-brightness
dwarf spheroidals, as expected, lie significantly below
this FP defined by the high surface brightness elliptical galaxies.
The conventional explanation is that these object are dominated
by dark matter in the inner regions.  With MOND, this is precisely the  
expectation for low surface brightness objects where a large
discrepancy is predicted and 
the size has disappeared as a parameter.  

\section{The universal Faber-Jackson relation}

\begin{figure}
\resizebox{\hsize}{!}{\includegraphics{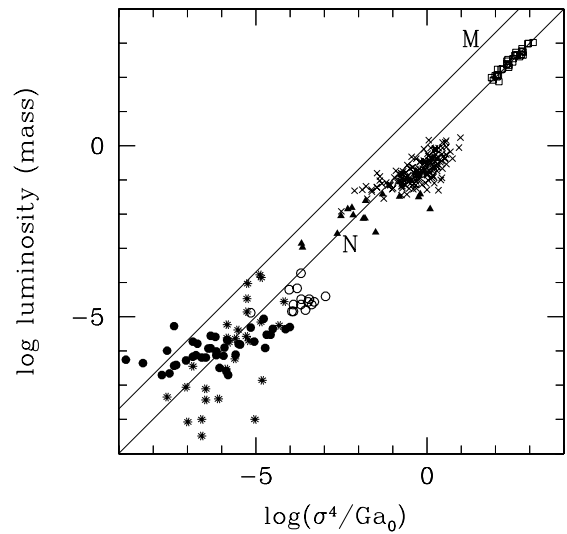}}
\caption[]{A log-log plot of the luminosity vs. the Faber-Jackson
quantity ${\sigma}^4/Ga_0$ in units of $10^{11}$ L$_\odot$ or
M$_\odot$.  The symbols and sources of data are the same as in
Fig.\ 3.  The parallel lines correspond to 
$M=c_2{\sigma_l}^4/Ga_0$ for the Jaffe model where again $\sigma_l$
is the intensity-weighted velocity dispersion within $\mathit{R_{eff}}$.  The upper
line is appropriate to a deep MOND object with $\eta=0.1$ (or to a
dark matter dominated homologous system) and the lower line to
a near Newtonian system with $\eta=10$ (or to a visible matter dominated
system).  Again, for the clusters of galaxies the mass of hot gas 
rather than luminosity is plotted.}
\label{}
\end{figure}

Fig.\ 4 is the FJ relation for the early-type galaxies shown in
Fig.\ 2 (from Cappellari et al. 2006).  Here the luminosity 
is plotted against $\log(\sigma^4/Ga_0)$;
the parallel lines are the relation
$$M=c_2 {\sigma_e}^4/Ga_0.\eqno(6)$$
The upper line (labelled M), with $c_2 = 20.25$, is relevant to the deep MOND 
limit, or equivalently the complete dominance of dark matter.  
The lower curve, with $c_2 = 1$, applies to highly Newtonian objects ($\eta = 10$)
or a Jaffe model with a very low dark matter content within $\mathit{R_{eff}}$.  
Objects with a 
fixed value of $\eta$ correspond to homologous structures having
a fixed mean surface density, $\Sigma$, within an effective radius.  
Constant surface density and the virial
theorem yield $\sigma^4\approx G^2M\Sigma$ or a $M\propto \sigma^4$ 
relation even for the Newtonian systems.
Large internal acceleration (high $\eta$) implies a large 
$\Sigma$ which implies that the Newtonian line, for the FJ relation,
lies to the right of the MOND line, opposite to the FP relation.
Again, the solid points are the same objects with luminosity multiplied
by M/L from the population synthesis models.  Although the scatter is
larger than for the FP relation, most of these points lie near
the the relation for Newtonian Jaffe models (low discrepancy).

The Faber-Jackson relation for the wider class of objects, globular clusters to clusters of galaxies, is shown in Fig. 5.
Most objects, ranging from Newtonian
to deep MOND, in so far as they can be approximated by the isotropic
Jaffe model, should lie in the range between the two parallel lines
if M/L is near unity.  The predicted scatter, 
due to a variation of $\eta$ only, is a factor of four
larger than for the FP relation.
As in Fig.\ 3, for the clusters of galaxies (open squares), 
it is the gas mass that
is plotted against the MOND parameter (${\sigma^4}/Ga_0$). As has been 
pointed out elsewhere (e.g., Sanders 2003) the cluster masses
calculated with MOND still require about three times more matter
than is directly observed in stars and hot gas.  Including
this ``missing mass" would move the
cluster points up by about 0.5 in the logarithm in
this plot.

Broadly speaking, this wide range of objects is
consistent with the FJ relation implied by MOND; that is to
say, there exists a Universal Faber-Jackson relation applying to
all near-isothermal pressure supported objects.  In particular,
the dwarf spheroidal systems, deep MOND objects, do not exhibit
such a large and systematic off-set from the mean relation as
they do in the case of the FP relation.  

\begin{figure}
\resizebox{\hsize}{!}{\includegraphics{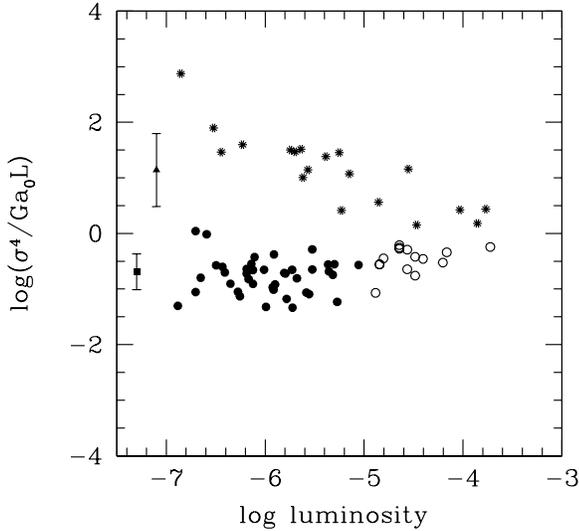}}
\caption[]{A log-log plot of $\sigma^2\mathit{R_{eff}}/GL$
vs. $L$ for dwarf spheroidals (starred points) , globular 
clusters (solid points) and
ultra-compact dwarfs (open points) over the same range of luminosity.
The triangular point with error bar is the mean and dispersion
of this quantity for the dwarf spheroidals and the square
point is the same for the globulars and UCDs taken together.
The result shows that the dwarf spheroidals significantly
deviate from the FP as defined by the more compact objects.}
\label{}
\end{figure}

\begin{figure}
\resizebox{\hsize}{!}{\includegraphics{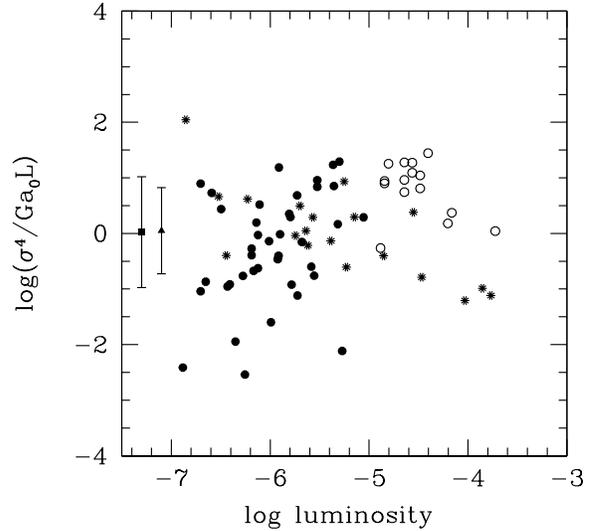}}
\caption[]{A log-log plot of $\sigma^4/Ga_0L$
vs. $L$ for dwarf spheroidals, globular clusters and
ultra-compact dwarfs over the same range of luminosity.
The triangular point with error bar is the mean and dispersion
of this quantity for the dwarf spheroidals and the square
point is the same for the globulars and UCDs taken together.
The result shows that there is no significant deviation
of the dwarf spheroidals from the FJ as defined by other classes
of pressure supported objects.}
\label{}
\end{figure}

The significance of this off-set is assessed in final two figures
which are alternative representations of the FP and FJ relations
over the range of luminosity where dwarf spheroidals overlap
the more compact objects.
Fig.\ 6 is a plot of the quantity $\log(\sigma^2 \mathit{R_{eff}}/
GL)$ against $\log(L)$ for dwarf spheroidals, globular clusters
and ultra-compact dwarfs from 
$10^4$ to $10^7$ L$_\odot$.  The plotted quantity is proportional to M/L as
estimated from the FP (virial theorem).  The points with error bars 
are the average and dispersion of this quantity -- the triangular
point for the dwarf spheroidals and the square point for
the globulars and UCDs taken together.  The dwarf spheroidals 
deviate by about two sigma from those systems lying on
the FP (this probably understates the significance
as the offset becomes more pronounced at lower luminosities).  
Fig.\ 7 is a plot of $\log(\sigma^4/Ga_0L)$ vs.
$\log(L)$ for these same systems.  This is proportional to M/L as
determined from the FJ relation (MOND).  Again the two points
with error bars are the mean and dispersion of this quantity
for the dwarf spheroidals (triangle) and separately the globulars
and UCDs (square).  Here we see that there is no significant
difference between these classes of objects; the dwarf spheroidals
lie on the universal Faber-Jackson relation defined by the more compact
pressure-supported objects.

\section{Conclusions}

Luminous elliptical galaxies, dwarf
elliptical galaxies, ultra-compact dwarfs, and globular star clusters
are high surface brightness objects.  High surface brightness
implies high internal acceleration; i.e., the gravitational acceleration
at $\mathit{R_{eff}}$ is roughly ten times larger than $a_0$,
the fundamental MOND acceleration parameter. 
This means that, in the context of MOND,
such objects should be described essentially by Newtonian dynamics within
an effective radius,   In that respect, it is not surprising that this
wide range of gravitationally bound pressure-supported structures
fall on a FP that reflects the Newtonian virial relation.
Of course there are variations: the globular clusters do
not lie precisely upon the same FP as the ellipticals, but
then the the data shown here is heterogeneous,
and the various classes of objects are probably not homologous.  
But the fact that the
dwarf spheroidal galaxies deviate significantly from this FP is consistent
with MOND. These are deep MOND low-surface-brightness objects
with a large discrepancy; in this deep 
MOND limit the length scale has 
dropped out as a parameter, and the mass is related only to
the velocity dispersion.

The high surface brightness objects defining the Newtonian FP must
deviate from an isothermal state, as they do in the case of the
bright ellipticals and the globular clusters;  both classes of
objects exhibit a line-of-sight velocity dispersion that 
declines with radius, at least within 1.5 $\mathit{R_{eff}}$.  At larger
radii, the radial profile of the $\sigma$ is highly dependent upon
the form of the anisotropy parameter $\beta$, but a continuing
decline is consistent with a trend toward more radial
orbits in the outer regions.

In order to produce a high surface density Newtonian object,
the required deviation from an isothermal state is relatively small.
Typically, the Jaffe models which are Newtonian
within an effective radius are equivalent to polytropes
of index 12 to 16; thus the density changes by many orders
of magnitude while the velocity dispersion changes by a 
factor of two.  In other words, the requirements for producing
objects which obey the Newtonian virial relation are
that they should have high internal accelerations ($\ge a_0$) and
be near-isothermal with a radially declining velocity dispersion.

One might argue that this again begs the question:  why are
ellipticals and globular clusters near-isothermal objects
with high internal accelerations?  Does 
this not push the existence of the FP back to contingencies of
structure formation?  To some extent, this is the case, but the
requirement of a near-isothermal state is rather mild. 
Violent relaxation, even in spherical collapse, typically
produces a near-isothermal virialized structure.  
For example, with modified dynamics just such objects with
high internal accelerations 
do result from the spherically symmetric
dissipationless collapse out of a medium initially expanding
with the Hubble flow (Sanders 2008).  
In non-spherical MONDian dissipationless collapse calculations,
the final virialized objects with low internal accelerations
($\le a_0$) are essentially isothermal, but also those with 
higher internal accelerations are near-isothermal
with a radial velocity dispersion which, compared with density, 
declines slowly with radius 
(Nipoti, Londrillo \& Ciotti 2007).  Viewed in this context,
the dwarf spheroidal galaxies become the anomalous objects
requiring an alternative formation scenario.

It is the FJ relation, in spite of its large scatter,
that emerges as the more fundamental and more universal scaling
relation for hot systems, embodying both high and low surface
brightness systems.  The amplitude of the relation is even
more significant than its slope because this is related directly
to the magnitude of $a_0$.  It is easily demonstrated 
that pure Newtonian
systems (obeying the Newtonian virial theorem) with a 
constant mean surface density will also fall on a
$M\propto \sigma^4$ relation.  The essential point is that 
$a_0$ sets this characteristic value of the
surface density ($a_0/G$) for all near-isothermal systems. 

The appearance of a universal
mass-velocity dispersion relation as an aspect of dynamics 
directly reflects the presence of this new dimensional constant in the
structure equation.
The fact that the magnitude of this constant ($\approx 10^{-8}$ cm/s$^2$)
is the same as that required by the scale of spiral galaxy
rotation curves (the Tully-Fisher relation) is one more powerful
indication that such a fundamental acceleration scale exists 
in the Universe and is operative in gravitational physics.

\vspace{5 mm}

I am grateful to Moti Milgrom and Joe Wolf for useful comments and to
Scott Trager for a critical reading of the paper.  This paper, in
content and presentation, has
benefitted from the comments of an anonymous referee.

\end{document}